\documentclass[aps,prb,twocolumn,showpacs,superscriptaddress,groupedaddress]{revtex4-1}

\usepackage{graphicx}  % needed for figures
\usepackage{dcolumn}   % needed for some tables
\usepackage{bm}        % for math
\usepackage{amssymb}   % for math
\usepackage{natbib}   % for biblio

\begin{document}

\title{Logic Gates with Bright Dissipative Polariton Solitons in Bragg-Cavity Systems}

\author{E.~Cancellieri}
\email[Corresponding author: ]{emiliano.cancellieri@gmail.com}
\affiliation{Department of Physics and Astronomy, University of Sheffield S3 7RH, UK}
\author{J. K.~Chana}
\affiliation{Department of Physics and Astronomy, University of Sheffield S3 7RH, UK}
\author{M.~Sich}
\affiliation{Department of Physics and Astronomy, University of Sheffield S3 7RH, UK}
\author{D. N.~Krizhanovskii}
\affiliation{Department of Physics and Astronomy, University of Sheffield S3 7RH, UK}
\author{M. S.~Skolnick}
\affiliation{Department of Physics and Astronomy, University of Sheffield S3 7RH, UK}
\author{D. M.~Whittaker}
\affiliation{Department of Physics and Astronomy, University of Sheffield S3 7RH, UK}
\date{\today}

\begin{abstract}
Optical solitons are an ideal platform for the implementation of communication lines, since they can be packed extremely close one to another without risking partial loss of the encoded information due to their interaction. On the other hand, soliton-soliton interactions are needed to implement computations and achieve all-optical information processing. Here we study how bright dissipative polariton solitons interact and exploit their interaction to implement AND and OR gates with state of the art technology. Moreover, we show that soliton-soliton interaction can be used to determine the sign of $\alpha_2$, the parameter describing the interaction between polaritons with opposite spin.
\end{abstract}

\maketitle 

\section{Introduction}
Solitons are self-reinforcing wavepackets that maintain their shape while propagating. Bright $dissipative$ solitons are a particular class of solitons typical of out-of-equilibrium systems. Differently from their $conservative$ counterpart, dissipative solitons are implemented by means of a continuous gain,  compensating the loss of particles from the system, and by means of a trigger, which generates a bright wavepacket on top of a much less intense background. This kind of solitons has been extensively studied in a wide range of systems and single and multiple solitons states as well as oscillating bound states have been predicted and demonstrated~\cite{dissipativesoliton,Pare1989228,multisoliton1997,PhysRevLett.85.748}. Surprisingly very little attention has been devoted to the study of interactions among bright dissipative solitons in dynamical systems~\cite{PhysRevE.53.6471,Akhmediev2005209} and, to the best of our knowledge, no attention has been devoted to the implementation of devices based on dissipative solitons, unlike for the conservative ones~\cite{Stegeman19111999}.

Since the first observation of strong light-matter coupling in a semiconductor micro-resonator~\cite{weisbuch1992} this system has been intensively studied for the implementation of a new generation of optical devices. In fact, the dual light-matter nature of polaritons, the particles emerging from the coherent strong coupling of cavity photons with quantum well excitons, provides significant advantages with respect to electronic as well as standard nonlinear optical systems. On the one hand the light component allows high propagation velocities and fast control and, on the other hand, the matter component guarantees strong nonlinearities, much stronger than in traditional optical systems with Kerr nonlinearities~\cite{tinkler2014}. For these reasons several theoretical proposals have been advanced for the implementation of neural networks, switches and logic gates~\cite{Liew2008,cancellieri2011,ortega2013}. And several experimental implementations such as switches, spin switches, resonant tunneling diodes and transistors have also been demonstrated~\cite{degiorgi,amo_switches,PhysRevLett.110.236601,ballarini2013,gao_savvidis}.

To implement fast and efficient polariton devices, however, the major problem of the long reset-time of the device needs to be solved. In fact, in all these approaches after each ``calculation'' one needs either to completely turn off the device or to wait hundreds of picoseconds to allow long-lived excitonic reservoirs to decay. A possible way to address this problem is to use red-detuned ultra-short Stark pulses~\cite{cancellieri2014} but with the major drawback that very high laser intensities are needed. Bright dissipative polariton solitons (BDPSs) can, in principle, solve this problem. In fact, BDPSs can be triggered with few ps-long pulses that do not excite excitonic reservoirs and, after the passage of a soliton, the system is left, by definition, in its OFF state.

In this work we study the interaction of dynamical BDPSs in circuits etched in planar Bragg-microcavities and show that AND and OR gates can be implemented by exploiting these interactions. Moreover, we show that by studying soliton-soliton interactions it is possible to evaluate the constant characterising the interaction of polaritons with different spins.

\begin{figure}
\centering
\includegraphics[width=1.0\linewidth]{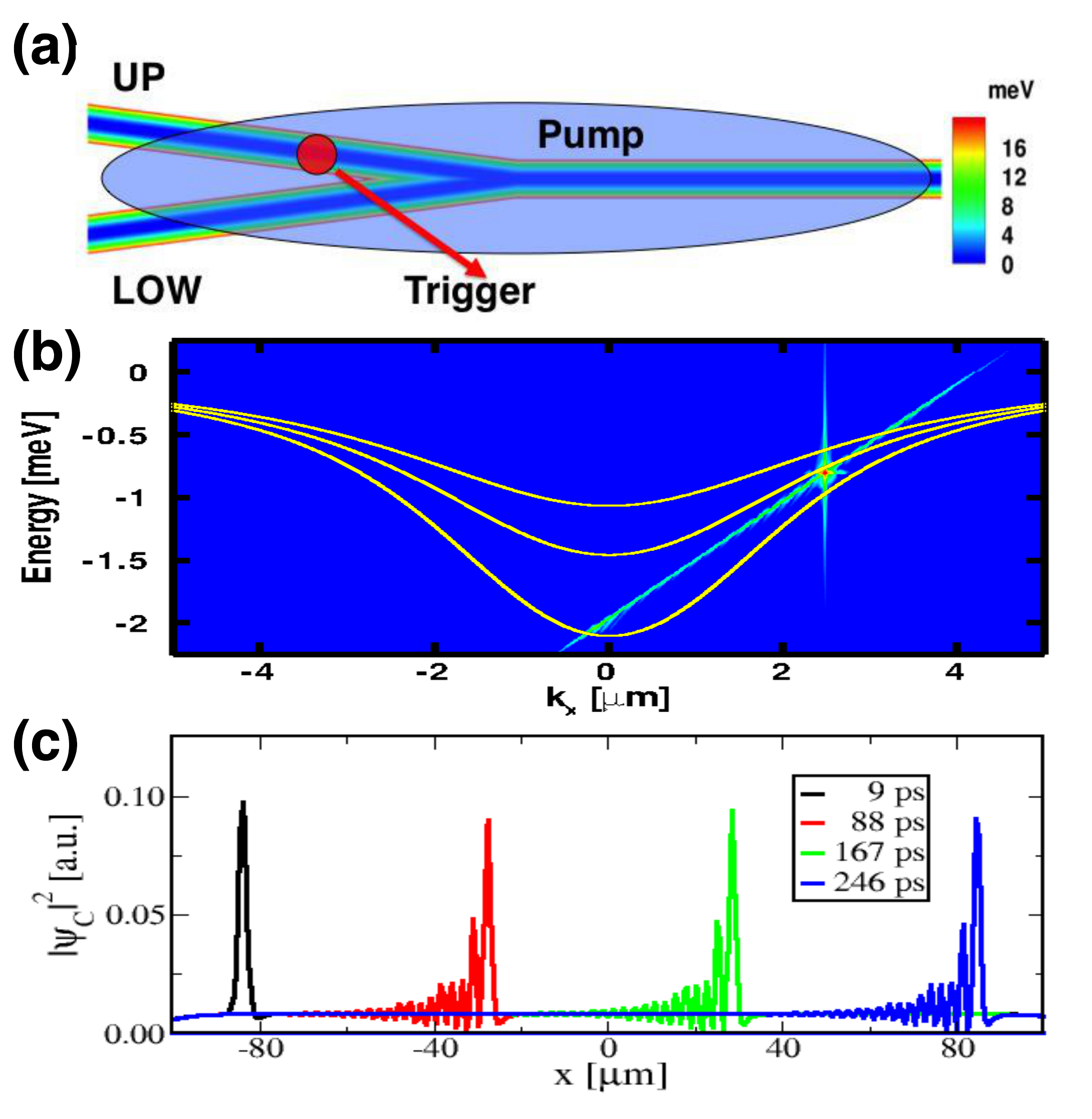}
\caption{\label{fig:soliton_wire}(Color online) (a) Schematic of the proposed device. (b) Spectra associated to $\psi^{ph}_+$ and lower polariton modes (yellow) induced by the harmonic potential $U^{\perp}_{\pm}(r)$. (c) Cuts of $|\psi^{ph}_+|^2$ along the centre of the wire at different times. The parameters are: $\hbar\gamma_{ph}=\hbar\gamma_{ex}=0.05$ meV, $\beta=0.0$ meV, $k_p=k_{pb_+}=2.5$ $\mu m^{-1}$, $\hbar\omega_p=\hbar\omega_{pb_+}=-0.8$ meV, $F^0_{\pm}=0.07$ meV$\mu m^{-1}$, $f^{pb}_{+}=0.5$ meV$\mu m^{-1}$, $\hbar\omega_{hp}=2.0$ meV (corresponding to wires 3 $\mu m$ wide).}
\end{figure}

The main requirements for the implementation of BDPSs, predicted and observed in planar 2-dimensional cavities~\cite{egorov2009,sich2011,sich2014}, are twofold: 1) the pump wavevector has to be above the point of inflection of the lower polariton branch, in order for the effective mass to be negative and compensate for the repulsive interactions between polaritons with the same spin component; 2) the pump has to be blue-detuned from the polariton branch in order to guarantee a bistable regime. In fact, BDPSs can be qualitatively interpreted as spatially localised excited regions of the modulationally unstable upper branch solution~\cite{sich2011}. The main advantage of BDPSs is that they can propagate for very large distances determined by the size of the pump. The main drawback, instead, is that being part of the pump state their energy, velocity and direction of propagation are set by the pump itself. Therefore, the propagation paths of two BDPSs can never intersect, making impossible the implementation of logic gates. To solve this issue we propose to inject BDPSs moving from left to right in circuits with the shape of a $Y$ [Fig.~\ref{fig:soliton_wire}(a)]. In this way the two converging arms act as filters allowing only the part of the pump parallel to them to enter in the cavity and forcing two BDPSs to meet at the junction.

The paper is structured as follows. In section \ref{Methods} we present the mean field approach used to describe the four components wavefuction (two exciton and two photon components, one for each spin/polarisation degree of freedom). Section \ref{Results} describes the main results of our research. First it will be demonstrated that dissipative solitons can propagate in straight wires and, making use of this result, the possible implementation of OR and AND gates will be addressed. The final part of section \ref{Results} will deal with the interaction of dissipative polariton solitons with opposite polarisation and with an analysis of the role played by the TE-TM splitting on the possible implementation of the gates. Conclusions and outlooks will be given in section \ref{Conclusion}.

\section{Methods}\label{Methods}
The system of a quantum well embedded in a semiconductor microcavity can be modelled by means of a four-component wave function where the spin up and spin down excitonic fields ($\psi^{ex}_{\pm}$) are strongly coupled to the two $\sigma_+$ and $\sigma_-$ circularly polarised photonic fields ($\psi^{ph}_{\pm}$) through the vacuum Rabi coupling $\hbar\Omega_R$ ($2.55$ meV in our case). The dynamics of the system can be modelled by means of the generalised Gross-Pitaevskii equation~\cite{PhysRevB.83.193305}:

\begin{eqnarray}
\label{eq:1}
i\hbar\partial_t\psi^{ph}_{\pm}
&=&
\left[\hbar\omega_{ph}({\bf k})+U_{\pm}-i\hbar\gamma_{ph}/2+\beta\left(ik_x\pm k_y\right)^2\right]\psi_{\pm}^{ph}
\nonumber
\\
&&
+F_{\pm}+\hbar\Omega_R\psi^{ex}_{\pm},
\nonumber
\\
i\hbar\partial_t\psi^{ex}_{\pm}
&=&
\left[\hbar\omega_{ex}({\bf k})-i\hbar\gamma_{ex}/2+\alpha_1|\psi^{ex}_{\pm}|^2+\alpha_2|\psi^{ex}_{\mp}|^2\right]\psi^{ex}_{\pm}
\nonumber
\\
&&
+\hbar\Omega_R\psi^{ph}_{\pm}.
\end{eqnarray}

Here $\omega_{ph}({\bf k})=\omega^0_{ph}+\frac{\hbar{\bf k}^2}{m_{ph}}$, with $m_{ph}=5.0\times 10^{-5}m_0$, is the dispersion of the confined photon mode and $m_0$ is the free electron mass. Since the exciton mass is much higher than $m_{ph}$ a flat exciton dispersion is taken ($\omega_{ex}({\bf k})=\omega_{ex}^0$). Throughout the paper the zero energy is set to the bare exciton frequency and exciton-photon detuning is taken to be equal to zero ($\omega_{ph}({\bf 0})=\omega_{ex}^0=0$). The terms $U_{\pm}$ describe wires etched in the top cavity-mirror. All the results shown here correspond to $U^{\perp}_{\pm}(r)=\frac{1}{2}m_{ph}\omega^2_{hp}d^2$, where $\omega_{hp}$ determines the potential strength and $d$ is the distance from the wires centre. In order to test the robustness of the proposed device we numerically confirmed the implementation OR and AND gates using potentials proportional to $d^3$ and to $d^4$. The parameters $\gamma_{ph},\gamma_{ex},\alpha_1$ and $\alpha_2$ respectively describe the photon and exciton decay rates, and the interaction between excitons with identical and opposite spin. For the sake of generality we rescale field densities and interaction constants to have $\alpha_1=1$ and $\alpha_2=-0.1$ \cite{PhysRevB.88.195302}. Finally, the terms proportional to $\beta$ describe the TE-TM splitting~\cite{PhysRevB.83.193305}, while the terms $F_{\pm}$ describe a continuous wave (CW) and a pulsed laser fields. Throughout the paper the CW terms will be taken equal to $F^0_{\pm}f({\bf r}) e^{i({\bf k}_p\cdot{\bf r}-\omega_p t)}$, where $f({\bf r})$ has a top-hat spatial profile, $\omega_p$ is the pump frequency and ${\bf k}_p=(k_p,0)$ is the pump wavevector. The pulsed terms have the same frequency and wavevector of the CW ones, intensity $f^{pb}_{\pm}$, and Gaussian profiles in space and time: $f^{pb}_{\pm}e^{i({\bf k}_p\cdot{\bf r}-\omega_p t)}e^{-x^2/2\sigma_{sp}^2-t^2/2\sigma_t^2}$, with $\sigma_t=1.5$ ps and $\sigma_{sp}=1$ $\mu m$.

\section{Results}\label{Results}
\subsection{Straight Wires}
We first consider the case of a single wire with zero TE-TM splitting and demonstrate the existence of solitonic solutions for harmonic potentials (see~\cite{Slavcheva:15} for super-Gaussian profiles). In analogy with~\cite{egorov2009,sich2011} we fix the CW-pump frequency and angle in order to be blue-detuned ($\Delta E=0.166$ meV) from the lower polariton dispersion at a wavevector above its point of inflection. Here, however, the relevant polariton dispersion is the one corresponding to the lowest confined mode of the harmonic trapping potential [Fig.~\ref{fig:soliton_wire}(b)]. Here and in the following we numerically simulate the time evolution of the system with a pump linearly polarised parallel to the wire and a trigger circularly polarised $\sigma_+$. Figure~\ref{fig:soliton_wire}(b) shows the typical solitonic linear dispersion while panel (c) shows 1-dimensional cuts along the propagation direction of the $\sigma_+$-polarised cavity emission ($\psi^{ph}_+$). After an initial transient time, in which the excited Gaussian-shaped wavepacket transforms into a solitary wave ($9-88$ ps), the profile does not change until the wavepacket exits the CW-pump, therefore demonstrating the existence of BDPS solutions (animation in supplementary material SVwire).

\begin{figure}
  \centering
  \includegraphics[width=1.00\linewidth]{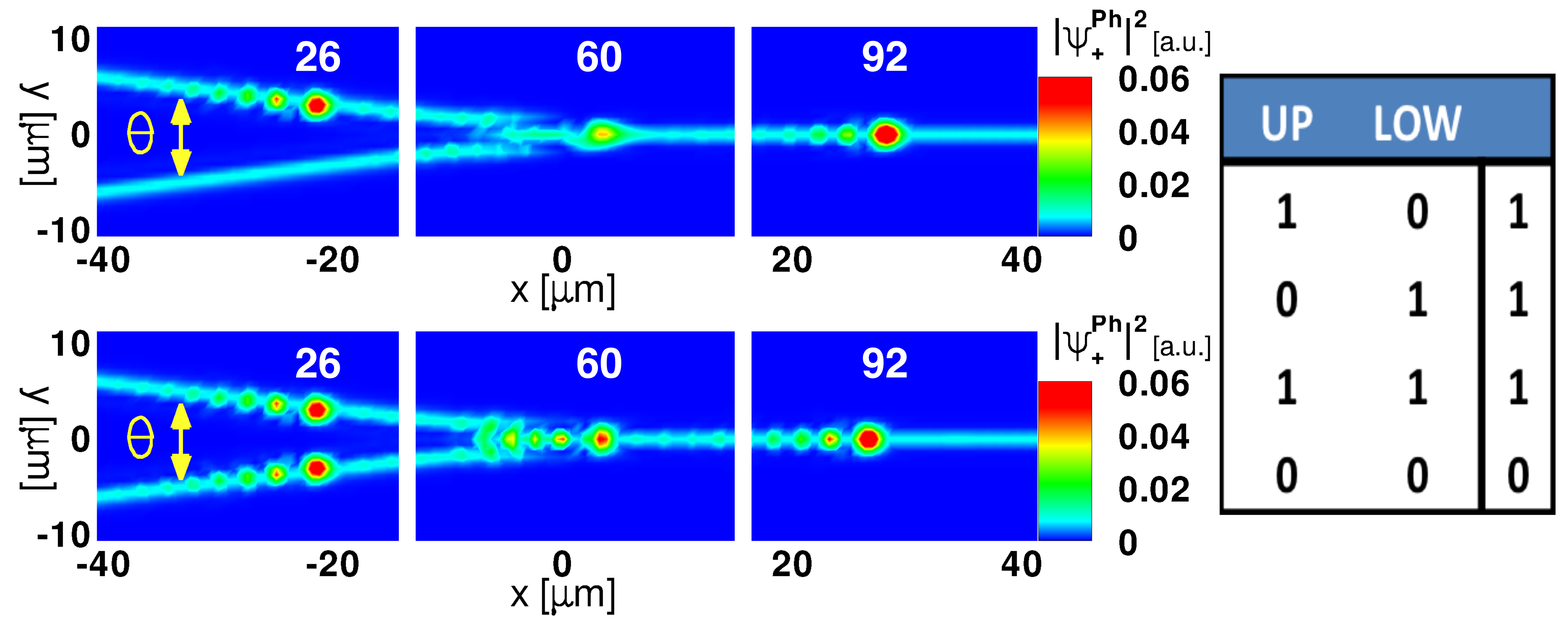}
  \caption{\label{fig:ORgate}(Color online) Or gate. In the logic table 1 represents a $\sigma_+$ polarised soliton and 0 the absence of it while UP and LOW indicate the upper or lower oblique arms. For the OR gate an output equal to 1 is expected every time there is at least a soliton in one of the inputs. The output 0 is expected only when no solitons enter the device. The top and bottom panels show the $\sigma_+$ cavity emission at three different times ($t=26, 60$ and $92$ps) for the first and third rows of the OR logic table. The parameters used here are the same as in Fig. \ref{fig:soliton_wire} and $\theta=16^{\circ}$.}
\end{figure}

\subsection{Logic gates: OR and AND}
Let us now address the case of $Y$ junctions still considering $\beta=0$. Since the CW-pump is set in order to inject polaritons moving from left to right, the region of the device downstream the junction is analogous to the case of the single wire. Therefore, considering the same CW-pump as before we are sure that BDPSs can propagate in this part of the device. The question is whether BDPS solutions exist in the two oblique wires. As already said, these wires essentially act as filters allowing only the components of the CW-pump parallel to them to enter into the microcavity. Therefore, inside the oblique wires the polariton wavevector is: $k_{\parallel}=k_pcos(\theta/2)$, where $\theta$ is the angle between the two wires. If $\theta$ is small enough so that $k_{\parallel}$ is above the point of inflection of the polariton dispersion soliton propagation is permitted. For our choice of the parameters ($\theta=16^{\circ}$) $k_{\parallel}=2.47$ $\mu m^{-1}$ is still above the point of inflection.

In order to demonstrate the implementation of logic gates we define the base of the binary logic as follows: $1$ corresponds to the presence of a BDPS polarised $\sigma_+$, and $0$ to the absence of it. Note that since we are considering $\beta=0$ and linearly polarised CW pumps, a BDPS polarised $\sigma_-$ as $1$ is an equivalent choice. Linearly polarised BDPSs, instead, cannot be used since the interaction of excitons with opposite spin components makes them unstable~\cite{sich2014}.

Figure~\ref{fig:ORgate} demonstrates the implementation of an OR gate. The top panel shows the $\sigma_+$ polarised emission from the cavity at different times: when the soliton is in the upper arm, when is just after the junction, and when is propagating downstream the junction. This panel shows that the first line of the OR logic table can be implemented using travelling BDPSs in $Y$ junctions (supplementary material SVOR1). What is it worth noticing is that at $t=60$ ps the polariton distribution does not have the typical BDPS-like distribution as at $t=26$ ps (a bright peak followed by a few less intense peaks). This can be understood by observing that in the junction the laser-lower polariton detuning ($\Delta E$) is not the same as in the wires, and therefore BDPSs entering the junction will need to adapt to the new conditions. For the specific set of parameters chosen here the bright wavepacket is able to pass through the junction, to enter into the downstream wire and, after few ps, to stabilise to the usual BDPS shape. However, if in the junction the deviation from the optimal condition for soliton formation is too large (i.e. $\Delta E$ is too big), the BDPS will disappear because the pump is too weak to sustain its excess population while in the junction.

The lower panel in figure~\ref{fig:ORgate} shows the implementation of the third line of the OR logic table. Here, two BDPSs simultaneously arriving at the junction transform into a single BDPS downstream the junction (supplementary material SVOR3). What is worth noticing here is that the two solitons merge into a single BDPS having lower intensity than the sum of the two initial ones. This is because the CW-pump  is fixed in order to sustain a single BDPS, and is too low to sustain the population of two BDPSs. Therefore, the population forming two BDPSs is bound to decrease down to the population of a single soliton. Finally, the second and fourth lines of the OR logic table are trivial if the first and third are satisfied.

\begin{figure}
  \centering
  \includegraphics[width=1.00\linewidth]{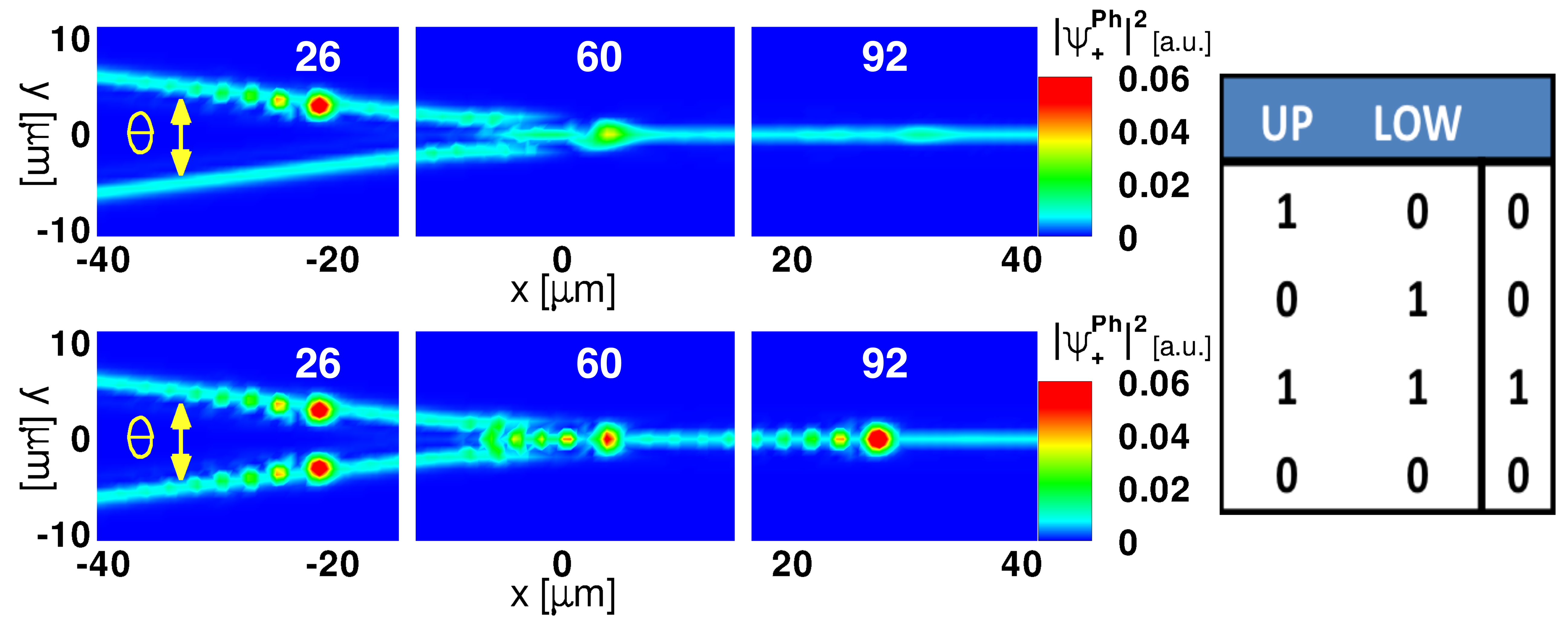}
  \caption{\label{fig:ANDgate}(Color online) AND gate. As in Fig.\ref{fig:ORgate} the top and bottom panels show $|\psi^{ph}_+|^2$ at $t=26, 60$ and $92$ps for the first and third rows of the logic table. The parameters here are the same as in Fig. \ref{fig:ORgate} and \ref{fig:soliton_wire} apart from the strength of the harmonic potential $\omega_{hp}=1.90$, which correspond to wider wires.}
\end{figure}

In order to implement an AND gate it is useful to recall what was just said about $\Delta E$ being, in the junction, too big to allow for the pump to sustain a BDPS. A possible way to implement an AND gate is to increase $\Delta E$ enough to forbid one soliton to pass, but keeping it small enough for two solitons to pass. Since we want to build a complex circuit we would like to use the same CW-pump over several devices and, therefore, we chose to increase $\Delta E$ by decreasing the value of $\omega_{hp}$, i.e. by increasing the wire size. Clearly, the change in the wire size between OR and AND gates will need to be smooth, in order to avoid reflections at the interface. Figure~\ref{fig:ANDgate} demonstrates the implementation of the AND gate in the case of wider wires (see animations in SVAND1 and SVAND3). In particular, the upper panel shows a BDPS not being able to cross the junction. This is because, in the junction the detuning $\Delta E$ is bigger with respect to the OR case and, therefore, it is more difficult for the pump to sustain the needed excess population. Instead, the lower panel shows two solitons simultaneously reaching the junction and a BDPS surviving downstream, therefore demonstrating the AND function. This is because even if many polaritons are lost while crossing the junction the amount of population due to the two initial BDPSs is enough to sustain a single BDPS.

Since the success of the operation relies on the sum of the population of the two BDPSs, their arrival at the junction must happen at the same time. For our choice of the parameters BDPSs travel at $\approx$0.7 $\mu m/ps$ and have a FWHM of about 1.5 $\mu m$. Therefore the needed precision over the time of their arrival is of the order of $\pm$1 ps (this value has been confirmed by our simulations). These values also allow determining the expected achievable repetition rate for the device: considering a minimal distance between solitons of four times the FWHM (6 $\mu m$), a repetition rate of the order of 100 GHz can be evaluated. Finally, in order to check the robustness of the proposed device we numerically confirmed that OR and AND gates can be implemented for a series of different parameters ($\omega_p$ and $k_p$ have been varied in a range of about 5$\%$ and 8$\%$ respectively).

\begin{figure}
  \centering
  \includegraphics[width=1.0\linewidth]{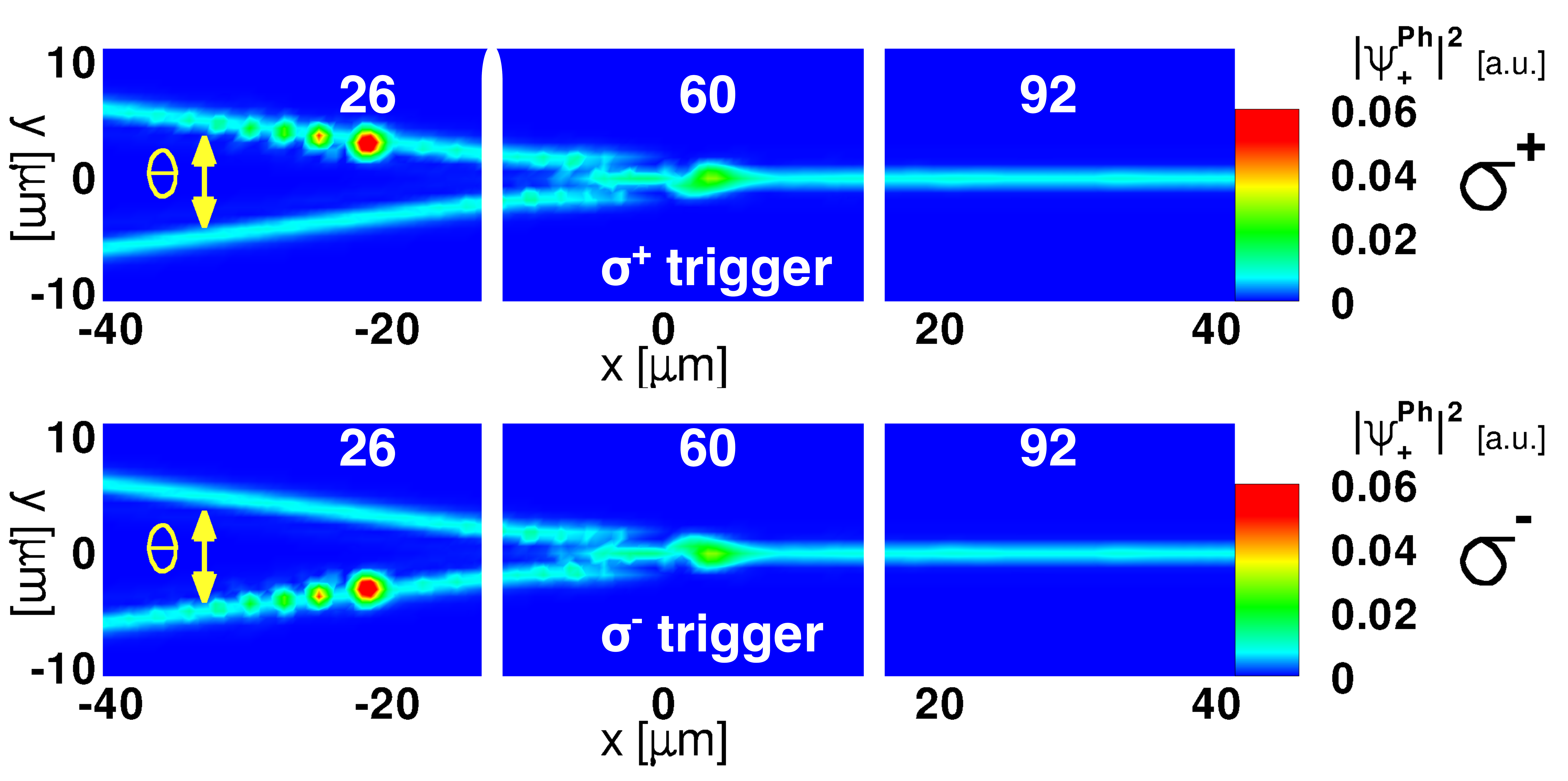}
  \caption{\label{fig:annihilation}(Color online) Interaction of $\sigma_+$ and $\sigma_-$ BDPSs in the case of $\alpha_2<0$. Top (bottom) panel: $\sigma_+$ ($\sigma_-$)-polarised cavity emission at $t=26, 60$ and $92$ps. A linearly polarised pump, as in Fig. \ref{fig:soliton_wire} and \ref{fig:ORgate}, is used to sustain $\sigma_+$ and a $\sigma_-$-polarised BDPSs respectively in the upper and lower arms. When two BDPSs meet at the junction they mutually annihilate. All parameters are as in Fig.~\ref{fig:soliton_wire} and~\ref{fig:ORgate}, except for the polarisation of the lower arm trigger.}
\end{figure}

\subsection{Interaction of $\sigma_+$ and $\sigma_-$  polarised solitons}
Let us now address the case of the interaction of BDPSs with different spin. From equation~(\ref{eq:1}) it can be seen that the effect due to the presence of a population of polaritons $\sigma_-$ ($\sigma_+$)-polarised is to shift the opposite-polarised polariton branch by an amount $\alpha_2|\psi^{ex}_+|$ ($\alpha_2|\psi^{ex}_-|$), with a positive or negative shift depending on the sign of $\alpha_2$. Figure~\ref{fig:annihilation} shows the same situation as in the lower panel of Fig.~\ref{fig:ORgate} but with a $\sigma_-$ polarised BDPS in the lower arm (this is possible since the CW-pump is linearly polarised). When the two BDPSs with different polarisation interact, they mutually annihilate (supplementary material SVALPHA2). In fact, since $\alpha_2$ is negative, the $\sigma_+$ polarised soliton red-shifts the $\sigma_-$ polarised polariton branch out of resonance with the CW-pump by an amount $\alpha_2|\psi^{ex}_+|$, therefore inducing the decay of the $\sigma_-$ polarised soliton. The same is valid for the $\sigma_-$ polarised soliton that induces the decay of the $\sigma_+$ polarised one. The situation would have been completely different for $\alpha_2>0$. In that case both BDPSs would have blue-shifted the polariton branch with opposite polarisation therefore helping the CW-pump to sustain the excess population.

\subsection{Effect of TE-TM splitting}
Finally let us comment on the effect of TE-TM splitting ($\beta\ne 0$). The case of BDPSs travelling in wires is not qualitatively different from the case of planar Bragg-microcavity. If the splitting induced by $|\beta|$ at the pump wavevector $k_p$ is small relatively to the detuning $\Delta E$, both $\sigma_+$ and $\sigma_-$ polarised BDPSs can be excited~\cite{sich2014}. This is understood in terms of an ``effective'' quenching of the TE-TM splitting. However, even if TE-TM splitting does not forbid, in principle, the implementation of the proposed logic gates, the condition $\Delta E>|\beta|k_p^2$ can be rather restrictive, specially for weak lateral confinement of the wires. In this case, in fact, the different polariton branches are quite close one to another and therefore the maximum allowed value for $\Delta E$ is small (we confirmed the feasibility of our devices for realistic values of $\beta=0.02$ meV$\mu m^2$~\cite{PhysRevB.83.193305}).

\section{Conclusion}\label{Conclusion}
In summary, we have theoretically demonstrated that high speed travelling BDPSs can cross junctions about 10 $\mu m$ wide and travel through Y-shaped devices. Moreover, we showed that two BDPSs meeting at the junction show effective interactions that can be exploited to implement OR and AND gates with repetition rates up to 100 GHz. This result is achieved in devices that allow BDPSs to move one towards another, overcoming the limitations in propagation direction typical of BDPSs in planar microcavities. Moreover, we showed that with the proposed devices it is possible to investigate the sign of $\alpha_2$, the parameter describing the interaction between excitons with opposite spin.

\acknowledgements
We acknowledge support by EPSRC grant EP/J007544, ERC Advanced grant EXCIPOL, and Leverhulme Trust.

\bibliography{mybiblio}

\end{document}